\documentstyle[12pt]{article}
\begin{document}
\vspace*{2cm}
\begin{center}
\begin{large}
{\bf On the interaction energy of 2D electron  FQHE
systems within the Chern-Simons approach}
\end{large}
\end{center}

\vspace{1cm}
\begin{center}
Piotr Sitko 
\vspace{0.3cm}

  Institute of Physics, 
Wroc\l{}aw University of Technology,\\ Wybrze\.ze Wyspia\'{n}skiego
 27, 50-370 Wroc\l{}aw, Poland.
\end{center}
 
 \vspace{1cm}
\begin{center}

Abstract

\end{center}
The interaction energy of  the two-dimensional electron system
in the region of fractional quantum Hall effect
is considered within the Chern-Simons composite fermion approach.
In the limit when Coulomb interaction is very small comparing
to the cyclotron energy
the RPA results are obtained
 for the fillings $\nu=1/3$, $1/5$, $2/3$, $2/5$, $3/7$  
 and compared with the exact diagonalization
results for small systems (extrapolated for infinite systems).
They show very poor agreement suggesting the need for looking
for alternative approaches.

\vspace{1cm}
keywords: fractional quantum Hall effect, composite fermions

\newpage
\noindent{\bf  1. Introduction}
\vspace{0.5cm}

One of the aims of 
 theoretical studies of fractional-quantum-Hall-effect 
(FQHE) systems \cite{Laughlin,Prange}
is to determine the value of the interaction energy
of the system at various fractional fillings 
of the lowest Landau level.
The exact numerical results can be found for few particle systems
\cite{Fano,Morf,Fano2} and
they agree very well with
the predictions
of the trial wave function approach of
Laughlin and Jain \cite{Laughlin,Jain,Jain2}
(Laughlin and Jain wave functions are proposed for the case when
Coulomb interaction is very small comparing to the cyclotron energy
and higher Landau levels can be omitted, it is also the case
in related numerical work).
The many-body theory for such systems is formulated within
the Chern-Simons gauge theory \cite{Lopez1,Halperin}
which introduces the gauge field  mapping
fermions into fermions (so-called composite fermions obtained
by attaching an even number of flux quanta to each electron).
There is no small parameter in the Chern-Simons  composite
fermion theory \cite{Halperin,Lopez2,Simon,Szewczenko} 
(in contrast to the case of anyons in the fermion limit
\cite{Chen}).
Nevertheless, the  Chern-Simons theory   
 gives good predictions of
the transport properties of the system \cite{Halperin,Lopez2}.  
However, in contrast to the trial wave function approach, 
the attempts to get the ground state energy of the system 
within the Chern-Simons theory are not very successful 
\cite{Szewczenko,Sitko}. In calculations of energy gaps
 modified approach have been used in which the relation between
finite size calculations and the Chern-Simons results is assumed
\cite{Halperin,Simon,He}. 
In this paper we calculate the interaction energy of the system
in the Chern-Simons approach within the RPA
for several values of the filling fraction
($\frac{1}{3},\frac{1}{5},
\frac{2}{3},\frac{2}{5},\frac{3}{7}$)
and compare them with the exact diagonalization results for 
few electron systems (extrapolated for infinite systems).

The composite fermion (CF) transformation consists in attaching
an even number ($2p$) of flux quanta to each electron.
Such point fluxes do not change the statistics of composite
particles, however, they allow to treat the 2D system of electrons
in a strong magnetic field in a close analogy to the treatment
in a weak magnetic field.
It is motivated by the mean field approach when
the sum of point fluxes is replaced by an average flux. 
The corresponding average  field is
$B^{Ch-S}=-2p\frac{hc}{e}\rho$ (fluxes opposite to the external flux,
$p$ is an integer, $\rho$ -- density) and
the effective field acting on electrons is reduced  to
$B^{eff}=B^{ex}+B^{Ch-S}$. In a close analogy to the quantum Hall
effect one predicts similar effect
when  $n$ Landau levels are completely filled in the effective field, i.e.
$B^{eff}=\frac{1}{n} \frac{hc}{e}\rho$. Hence,
$B^{ex}=\frac{2pn+1}{n} \frac{hc}{e}\rho$, which means that 
 the "real" lowest Landau level
is filled in the fraction $\nu=\frac{n}{2pn+1}$
($\nu=\frac{n}{2pn-1}$ when the effective field is opposite 
to the external one).
It is interesting to notice that the Laughlin fractions (of the form
$1/m$, $m$ -- odd) can be represented in two different ways.
We can add that the Chern-Simons theory results should be independent
of the actual Chern-Simons parameter $2p$ (if even) \cite{Fradkin}.
Hence, we can use the value of $p$ which is most suitable in a given
problem.
In practice, we use the value of $p$ which gives fully filled Landau levels
in the effective field (treatment of such systems is well known).
Nevertheless, whatever value of $p$  is taken the results should be the
same.

The Hamiltonian of the two-dimensional system of 
electrons in an external magnetic field
\begin{displaymath}
H=\int d^{2}{\bf r} \Psi^{+}({\bf r})
\frac{1}{2m}
({\bf p}+\frac{e}{c}{\bf A}^{ex}({\bf r}))^{2}
\Psi({\bf r}) 
\end{displaymath}
\begin{equation}
+\frac{1}{2}\int\int d^{2}{\bf r}d^{2}{\bf
r}'\Psi^{+}({\bf r})\Psi^{+}({\bf r}')\frac{e^{2}}
{\epsilon|{\bf r}-{\bf r}'|}
\Psi({\bf r})\Psi({\bf r}')\;
\end{equation}
($\epsilon$ -- dielectric constant) can be rewritten in the following way:
\begin{displaymath}
H=\int d^{2}{\bf r} \Psi^{+}({\bf r})
\frac{1}{2m}
({\bf p}+\frac{e}{c}{\bf A}^{ex}({\bf r})
+\frac{e}{c}{\bf A}^{Ch-S}({\bf r}))^{2}
\Psi({\bf r}) +
\end{displaymath}
\begin{equation}
+\frac{1}{2}\int\int d^{2}{\bf r}d^{2}{\bf
r}'\Psi^{+}({\bf r})\Psi^{+}({\bf r}')\frac{e^{2}}
{\epsilon|{\bf r}-{\bf r}'|}
\Psi({\bf r})\Psi({\bf r}')\;
\end{equation}
where
\begin{equation}
A^{Ch-S}_{\alpha}({\bf r}) = -2p
\frac{\hbar c}{e}\int d^{2}{\bf r}\epsilon_{\alpha
\beta}\frac{({\bf r}-{\bf r}')_{\beta}}{|{\bf r}-{\bf
r}'|^{2}}\rho(\bf r)   \;,
\end{equation}
 $\rho({\bf r})=\Psi^{+}({\bf r})\Psi({\bf r})$.  
The Hamiltonian $H$ can be separated into two parts:
$H=H_{0}+H_{int}$ 
where 
\begin{equation}
H_{0}=\int d^{2}{\bf r} \Psi^{+}({\bf r})
\frac{1}{2m}
({\bf p}+\frac{e}{c}{\bf A}^{ef}({\bf r}))^{2}
\Psi({\bf r}) 
\end{equation}
is  treated as the
unperturbed term ($B^{eff}=\nabla\times {\bf
A}^{ef}=\nabla\times ({\bf A}^{ex}+{\bf \bar{A}}^{Ch-S})=
B^{ex}+B^{Ch-S}$, $B^{Ch-S}$ is found by averaging point fluxes
-- putting the average density $\rho$ in (3)). 
$H_{int}$ is the {\it interaction} Hamiltonian	 \cite{Chen,Dai}:
\begin{equation}
H_{int}=\frac{1}{2}\int\int d^{2}{\bf r}d^{2}{\bf
r}'\Psi^{+}({\bf r})\Psi^{+}({\bf r}')\frac{e^{2}}{\epsilon|{\bf r}-{\bf r}'|}
\Psi({\bf r})\Psi({\bf r}')+H_{1}+H_{2} \; ,
\end{equation}
where
\begin{equation}
H_{1}=-2p\frac{\hbar}{m}\int\int d^{2}{\bf r}d^{2}{\bf
r}'\Psi^{+}({\bf r})(p_{\alpha}+ 
\frac{e}{c}A_{\alpha}^{ef})\Psi({\bf r})
\epsilon_{\alpha
\beta}\frac{({\bf r}-{\bf r}')_{\beta}}{|{\bf r}-{\bf
r}'|^{2}}(\rho({\bf r}')-\rho),
\end{equation}
\begin{equation}
H_{2}=(2p)^{2}\frac{\hbar^{2}}{2m}\int\int\int d^{2}{\bf r}d^{2}{\bf r}'d^{2}{\bf r}''
\rho({\bf r})\frac{({\bf r}-{\bf r}')}{|{\bf r}-{\bf
r}'|^{2}}\frac{({\bf r}-{\bf r}'')}{|{\bf r}-{\bf
r}''|^{2}}(\rho({\bf r}')-\rho)(\rho({\bf r}'')-\rho).
\end{equation}
In this paper we consider the case
when $B^{eff}=\frac{1}{n} \frac{hc}{e}\rho$, i.e.  in the unperturbed
state one has $n$ completely filled Landau levels (the effective filling
$\nu^{*}=n$).
The first step in calculating the ground state interaction
energy is the Hartree-Fock
approximation.
Considering the Coulomb interaction one finds the Hartree-Fock (H-F)
contribution to be
(we assume the presence of the positive background):
\begin{equation}
E^{H-F}=-\frac{N}{2n} \frac{e^{2}}{\epsilon a_{0}^{eff}}
\int_{0}^{\infty}[\sum_{k=0}^{n-1} L_{k}^{0}(\frac{1}{2}r^{2})]^{2}
\exp{(-\frac{1}{2}r^{2})}dr\; 
\end{equation}
where $a_{0}^{eff}=\sqrt{\frac{\hbar c}{eB^{eff}}}$ is the effective
magnetic length 
($a_{0}^{eff}=\sqrt{\frac{B^{ex}}{B^{eff}}}a_{0}^{ex}=
\sqrt{2pn\pm 1}a_{0}^{ex}$,  $a_{0}^{ex}=\sqrt{\frac{\hbar c}{eB^{ex}}}$),
$L_{l}^{m}$ -- Laguerre polynomials, $N$ -- number of particles.
The H-F results are presented in Table I for
several filling fractions and compared with "exact" results
(exact diagonalization results extrapolated for infinite systems).
The difference between "exact" and Hartree-Fock
results (correlation energy) increases with the decrease of the fraction,
for the $1/3$ state  is 
 of order of $10$\% (of the exact value).
 It seems that a higher order approximation, eg.  the RPA,
 will give a better agreement.
In the following we consider the correlation energy
within the RPA, assuming that the separation between Landau
levels is much larger than Coulomb interaction between particles
(as it is the case in exact diagonalization methods we refer to).

\newpage
\vspace{0.5cm}
\noindent{\bf 2. Correlation energy}
\vspace{0.5cm}

The correlation energy can be defined as follows:
\begin{equation}
E_{c}=\int_{0}^{1}\frac{d\lambda}{\lambda}(<\lambda H_{int}>_{\lambda}
-<\lambda H_{int}>_{0})\; .
\end{equation}
The expression for the correlation energy in the RPA
(three-body
contributions are omitted) has the form \cite{Hanna,Szewczenko}
\begin{equation}
E_{c}^{RPA}	= -\frac{1}{2}\hbar L^{2}\int\frac{d{\bf q}}{(2\pi)^{2}}
\int_{0}^{\infty}\frac{d\omega}{\pi}\int_{0}^{1}\frac{d\lambda}{\lambda}
{\rm Im\; tr} (\lambda V(q))[D_{\lambda}^{RPA}({\bf q},\omega)-D_{0}({\bf
q},\omega )]
\end{equation}
where $D_{\lambda}^{RPA}$ is the correlation function of
effective field currents ($L^{2}$ is the area of the system):
\begin{equation}
D_{\mu\nu}^{RPA}({\bf r} t,{\bf r'} t')=
-\frac{i}{\hbar}<T[j^{\mu}({\bf r} t),j^{\nu}({\bf r'} t')]> 
\end{equation}
given within the random-phase approximation (with the coupling
constant $\lambda$):
\begin{equation}
D_{\lambda}^{RPA}
({\bf q} ,\omega)=[I-\lambda D_{0}
({\bf q} ,\omega)V({\bf q})]^{-1}D_{0}
({\bf q} ,\omega).
\end{equation}
The current densities are defined as:
\begin{equation}
\bf j \rm ({\bf r})=\frac{1}{2m}\sum_{j}\left\{{\bf P}_{j}
+\frac{e}{c}{\bf A}_{j}^{ef},\delta({\bf r} -{\bf r}
_{j})\right\} 
\end{equation}
where braces denote an anticommutator,
  {\bf j} is the vector part
of $j^{\mu}$ with $\mu=0,x,y$. We define
$j^{0}$ as density fluctuations:
$j^{0}=\sum_{j}\delta
({\bf r} -{\bf r} _{j})-\rho$.
The interaction matrix $V$ is obtained from
the Hamiltonian $H_{int}$ (dropping out three-body terms).
We choose ${\bf q}=q\hat{\bf x}$ and the Coulomb gauge 
 which reduces the problem to $2\times 2$ $D^{RPA}_{\mu\nu}$ 
 matrix ($\mu=0,y$) \cite{Halperin}.
Taking  $\omega_{c}^{eff}=\frac{eB^{eff}}{cm}$ and
$a_{0}^{eff}$
to be 
frequency and length units, 
respectively one finds ($\hbar=1$):	
\begin{equation}
V({\bf q} )=\left(
\begin{array}{cc}
v(q)&0\\
0&0
\end{array}\right)+ 
\frac{4p\pi}{q^{2}}\left(
\begin{array}{cc}
2pn&-iq\\
iq&0
\end{array}\right)  
\end{equation}
where $v(q)$ is the Fourier transform
of the Coulomb potential ($v(q)=\frac{2\pi e^{2}}{\epsilon q}$ in standard
units).

Let us assume the correspondence between 
the two energy scales, one related to the separation
between Landau levels ($\hbar\omega_{c}^{ex}$), one to the strength of
the Coulomb interaction ($\frac{e^{2}}{\epsilon a_{0}^{ex}}$).
We introduce the dimensionless parameter:
\begin{equation}
r_{s}=\frac{e^{2}}{\epsilon a^{eff}_{0}}\cdot \frac{1}{\hbar\omega^{eff}_{c}}
\end{equation}
which shows the strength of the Coulomb interaction with respect
to the separation between effective Landau levels
($r_{s}^{ex}=\frac{e^{2}}{\epsilon a_{0}^{ex}}\cdot
\frac{1}{\hbar\omega_{c}^{ex}}=\frac{1}{\sqrt{2pn\pm 1}}r_{s}$).
If one considers the system of electrons,
the limit 
$\frac{e^{2}}{\epsilon a_{0}^{ex}}\cdot
\frac{1}{\hbar\omega_{c}^{ex}}\longrightarrow 0$ 
corresponds to the case when particle-hole excitations 
(electron excited into a higher Landau level) are negligible
(hence, in exact diagonalization studies higher Landau levels can be
omitted).
When applying the Chern-Simons picture, however, gauge interactions
are always of order of $\hbar\omega_{c}^{ex}$, and particle-hole excitations
(CF excited into an empty Landau level and a CF hole in a filled level)
have to be considered.
We have
\begin{equation}
V({\bf q} )= 
\frac{4p\pi}{q^{2}}\left(
\begin{array}{cc}
2pn(1+\frac{q}{(2pn)^{2}}nr_{s})&-iq\\
iq&0
\end{array}\right)  
\end{equation}
The correlation function $D_{0}$ is \cite{Jacak}: 
\begin{equation}
D_{0}
({\bf q} ,\omega)=\frac{n}{2\pi}\left(
\begin{array}{cc}
q^{2}\Sigma_{0}&-iq\Sigma_{1}\\
iq\Sigma_{1}&\Sigma_{2}
\end{array}\right)
\end{equation}
where
\begin{displaymath}
\Sigma_{j}=\frac{e^{-x}}{n}\sum_{m=n}^{\infty}\sum_{l=0}^{n-1}
\frac{m-l}{(\omega)^{2}-(m-l-i\eta)^{2}}
\frac{l!}{m!}x^{m-l-1}[L_{l}^{m-l}(x)]^{2-j}
\end{displaymath}
\begin{equation}
\times [(m-l-x)L_{l}^{m-l}(x)+2x\frac{dL_{l}^{m-l}(x)}{dx}]^{j}
\end{equation} 
and $x=\frac{q^{2}}{2}$.
Then one obtains:
\begin{equation}
D^{RPA}
({\bf q} ,\omega)=\frac{n}{2\pi det}\left(
\begin{array}{cc}
q^{2}\Sigma_{0}&-iq\Sigma_{s}\\
iq\Sigma_{s}&\Sigma_{p}
\end{array}\right)
\end{equation}
where
 $det=det(I-D^{0}V)=(1-2pn\Sigma_{1})^{2}-(2pn)^{2}\Sigma_{0}(1+\Sigma_{2})
 -nr_{s}q\Sigma_{0}$,\\
 $\Sigma_{s}=\Sigma_{1}-2pn\Sigma_{1}^{2}+2pn\Sigma_{0}\Sigma_{2}$,
 $\Sigma_{p}=(2pn)^{2}\Sigma_{1}^{2}+\Sigma_{2}-(2pn)^{2}\Sigma_{0}
 \Sigma_{2}+qnr_{s}(\Sigma_{1}^{2}-\Sigma_{0}\Sigma_{2})$.\\
Collective modes are determined by the poles of the correlation
function $D^{RPA}$.
In Figures 1-2  we plot the results for $\nu=1$ (the direct
result and the $p=1$  CF approach result) and similar results
for
 $\nu=1/3$ are presented in Figures 3-4, in Figure 5 the $3/7$ case is
presented.

The RPA correlation energy will be found using the dispersion relation
of collective modes.
In units of $\frac{e^{2}}{\epsilon a_{0}^{eff}}$
the correlation energy can be
expressed as follows \cite{Hanna} 
($\frac{e^{2}}{\epsilon a_{0}^{eff}}=r_{s}\hbar\omega_{c}^{eff}$ ):
\begin{equation}
E_{c}^{RPA}	= \frac{N}{2nr_{s}}\int_{0}^{\infty} q dq
\int_{0}^{\infty}\frac{d\omega}{\pi}
{\rm Im} (\ln{det}+{\rm tr}(V(q)D_{0}))
\end{equation}
which  equals
\begin{equation}
E_{c}^{RPA}	= \frac{N}{2nr_{s}}\int_{0}^{\infty} q dq
\int_{0}^{\infty}\frac{d\omega}{\pi}
{\rm Im} (\ln{det}+2pn(2pn\Sigma_{0}+2\Sigma_{1})+nr_{s}q\Sigma_{0}).
\end{equation}
We have:
\begin{equation}
\int_{0}^{\infty}dx\int_{0}^{\infty}\frac{d\omega}{\pi}
{\rm Im}\Sigma_{0}(x,\omega)=-\frac{1}{2}\sum_{m=1}^{\infty}\frac{1}{m}
+\frac{1}{2}(S_{n}-1)
\end{equation}
($S_{n}=\sum_{j=1}^{n}\frac{1}{j}$). This term is divergent but
combined with other terms in the integral (21) has to give
a finite value. Additionally
\begin{equation}
\int_{0}^{\infty}dx\int_{0}^{\infty}\frac{d\omega}{\pi}
{\rm Im}\Sigma_{1}(x,\omega)=0.
\end{equation}
The last integral in (21)
\begin{equation}
\label{add}
\frac{N}{2}
\int_{0}^{\infty}q^{2}dq\int_{0}^{\infty}\frac{d\omega}{\pi} {\rm Im}\Sigma_{0}.
\end{equation}
will be calculated separately for different $\nu^{*}=n$.

\vspace{0.5cm}
\noindent{\bf 3. Results}
\vspace{0.5cm}

To calculate the correlation energy (21) one needs to know the zeros
of the determinant $det$ (collective modes). For $n=1$ 
(the effective filling $\nu^{*}=1$ -- Laughlin fractions)
we have an infinite set of modes with shortwavelength
behaviour like $\omega_{m}(q\rightarrow\infty)=m$ -- Figs.1-4 \cite{Lopez2}. 
It can
be shown that \cite{Hanna}:
\begin{equation}
\int_{0}^{\infty}\frac{d\omega}{\pi}{\rm Im} \ln{det}=\sum_{m=1}^{\infty}
(\omega_{m}-m)=\sum_{m=1}^{\infty}\Delta\omega_{m}
\end{equation}
The integral (\ref{add}) equals:
\begin{equation}
\int_{0}^{\infty}q^{2}dq\int_{0}^{\infty}\frac{d\omega}{\pi} {\rm Im}\Sigma_{0}=
\int_{0}^{\infty}\sqrt{2x}\sum_{m=1}^{\infty}\frac{e^{-x}x^{m-1}}{m!}=
 \sqrt{2}\int_{0}^{\infty}\sqrt{x}(1-e^{-x})\; .
\end{equation}
Again, this integral is divergent, but combined with (25) and (22) 
 will give a finite result. We write:
\begin{equation}
\int_{0}^{\infty}q^{2}dq\int_{0}^{\infty}\frac{d\omega}{\pi} {\rm Im}
\Sigma_{0}=
-\frac{1}{2}\sqrt{2\pi}\sum_{m=1}^{\infty}\frac{1}{2^{m}m!}
\prod_{i=1}^{m}(2i-1)\;
\end{equation}
and then (in units of $\frac{e^{2}}{\epsilon a_{0}^{eff}}$)
\begin{equation}
\frac{E_{c}^{RPA}}{N}	= \frac{1}{2r_{s}}\sum_{m=1}^{\infty}
(\int \Delta\omega_{m}(x)dx -(2p)^{2}\frac{1}{2m}
-\frac{1}{2}r_{s}\sqrt{2\pi}\prod_{i=1}^{m}\frac{2i-1}{2i})
\; .
\end{equation}
In the case when $n\ge 2$
 every root (of collective modes) higher than the first is splitted
into two \cite{Hanna} (for $m>1$,
$\omega_{m}^{-}(q\rightarrow\infty)=m=\omega_{m}^{+}(q\rightarrow\infty)$
-- Figure 5).
One has 
\begin{equation}
\int_{0}^{\infty}\frac{d\omega}{\pi}{\rm Im}
\ln{det}=\Delta\omega_{1}+\sum_{m=2}^{\infty}
(\Delta\omega_{m}^{-}+\Delta\omega_{m}^{+})
\end{equation}
and the correlation energy for $n=2$ is  given by
(in units of $\frac{e^{2}}{\epsilon a_{0}^{eff}}$)
\begin{displaymath}
\frac{E_{c}^{RPA}}{N}	= \frac{1}{4r_{s}}(\int \Delta\omega_{1}(x)dx
-8p^{2})
+\frac{1}{4r_{s}}\sum_{m=2}^{\infty}
[\int (\Delta\omega_{m}^{-}(x)+\Delta\omega_{m}^{+}(x))dx
-(4p)^{2}\frac{1}{2m}]+\frac{p^{2}}{r_{s}}
\end{displaymath}
\begin{equation}
-\frac{1}{8}
\sqrt{2\pi}\sum_{m=1}^{\infty}
[(1-\delta_{m1})-m
+\frac{(2m+3)(2m+1)}{4(m+1)}]\prod_{i=1}^{m}\frac{2i-1}{2i}.
\end{equation}
We have also found the  expression for $E_{c}^{RPA}$ 
for $n=3$ (applied for $\nu=3/7$).

The value of interaction energy related
to Coulomb interaction
(in the limit of\linebreak $r_{s}\longrightarrow 0$) is:
\begin{equation}
\frac{\Delta E_{c}^{RPA}}{N}=\lim_{r_{s}\longrightarrow 0}
(\frac{E_{c}^{RPA}(r_{s})}{N}
-\frac{E_{c}^{RPA}(NC)}{N})\; ,
\end{equation}
$NC$ stands for the case with no Coulomb interaction \cite{Sitko}.
For $n=1$ one has
\begin{equation}
\frac{\Delta E_{c}^{RPA}}{N}
=\lim_{r_{s}\longrightarrow 0}\frac{1}{2r_{s}}\sum_{m=1}^{\infty}
(\int (\omega_{m}^{r_{s}}(x)-\omega_{m}^{NC}(x))dx 
-\frac{1}{2}r_{s}\sqrt{2\pi}\prod_{i=1}^{m}\frac{2i-1}{2i})
\; .
\end{equation}
The main problem in calculating (32) is the calculation of the integrals
and the convergence in summation over $m$.
The integrals in (32) (and similar for $n=2$ and $n=3$)
have been calculated numerically using $k$-point
Gauss-Laquerre integration. It was verified that the
summation over $m$ converges well and the sums have been truncated at
$2k$ terms.
In order to find the limit $r_{s}\longrightarrow 0$ we considered
small values of $r_{s}$. It appears that
for $r_{s}$ of order of $10^{-4}-10^{-5}$
the RPA energies
become practically independent of $r_{s}$ 
and for that range the results are given in Tables II-IV.

In Table II we present the RPA results in respective units of
$\frac{e^{2}}{\epsilon a_{0}^{eff}}$. 
It can be seen that the results for the same value
of $p$ and $|\nu^{*}|$ are very close one to the other. 
They are not the same for the same
filling (as they should be).
We observe rather strong dependence on $p$ 
(at a given effective filling $\nu^{*}$).

In Table III and Table IV the RPA results  are compared
with the exact diagonalization results.
The best agreement with the "exact" values is found
for the series $3/7$, $2/5$, $1/3$ (fractions going down
from $1/2$). 
The $1/3$ state gives the best result, the difference between the RPA
and the exact values is of order of 20\% (of the exact result).

An interesting example  is the system of electrons at the 
real filling $\nu=1$.
In the limit $r_{s}^{ex}\longrightarrow 0$  the RPA correlation energy
is zero.
The qualitative difference is found within the RPA for the CF approach.
Performing the CF transformation one finds the system
at the effective filling $\nu^{*}=-1$ (the effective field
is opposite to the external one).
We plot collective modes for the two descriptions in Figures 1-2.
Calculating the RPA result (as it is described above)
we find a finite CF result 
for the correlation energy ($r_{s}\longrightarrow
0$).

Similar situation can be found at the filling $\nu=1/3$.
Then the effective filling is  $\nu^{*}=1$ ($p=1$) or $\nu^{*}=-1$ ($p=2$)
and the Hartree-Fock contributions are the same in the two descriptions.
The RPA collective modes spectra look very similar -- Figures 3-4
(the agreement is exact at $q\rightarrow 0$)
but the values of interaction 
energy differ a lot (Table III and Table IV).

In summary the agreement between the RPA interaction energies
and the exact results are far from the expected one
and the analysis needs an extension by including
three-body contributions (three-body density-density correlation
function \cite{Hanna}) which seems to be very complicated. 
An alternative approach may be obtained
within the new  formalism developed by Shankar and Murthy \cite{Shankar}
(which has a direct relation with the Laughlin trial wave function). 
In Ref. \cite{Murthy1} they showed how to calculate the energy gaps
and their results agree reasonably with numerical results \cite{Murthy2}.

\vspace{0.5cm}
\noindent{\bf 4. Conclusions}
\vspace{0.5cm}

The values of Coulomb interaction energies for the 2D electron system
in the region of FQHE are calculated within the Chern-Simons theory
in the RPA for several
fractional fillings.
The results are obtained in the limit
$(\frac{e^{2}}{\epsilon a_{0}^{ex}})(\frac{1}{\hbar\omega_{c}^{ex}})
\longrightarrow 0$ (i. e. , when Coulomb interaction is very small
comparing to the separation between Landau levels)
and compared with the exact diagonalization results (the 
 results for few particle systems  extrapolated
for infinite systems).
The best agreement is found for 
fractions going down from $1/2$ ($3/7$, $2/5$, $1/3$),
for the best $1/3$ result the difference between
the RPA and the exact results is of order of $20$\% 
(of the exact value).
A qualitative difference is obtained for $\nu=1$ $p=1$ CF description
when a finite RPA correlation energy is found.
Also the result for $\nu=1/3$ $p=2$ is very different
from $\nu=1/3$ result obtained for $p=1$.
Our analysis needs an extension to a higher order approximation
 including  three-body contributions.
An alternative approach may be obtained 
within the new formalism developed by Shankar and Murthy 
\cite{Shankar}, their results for energy gaps \cite{Murthy1}
are in reasonable
agreement with numerical results \cite{Murthy2}.


\newpage

\noindent  
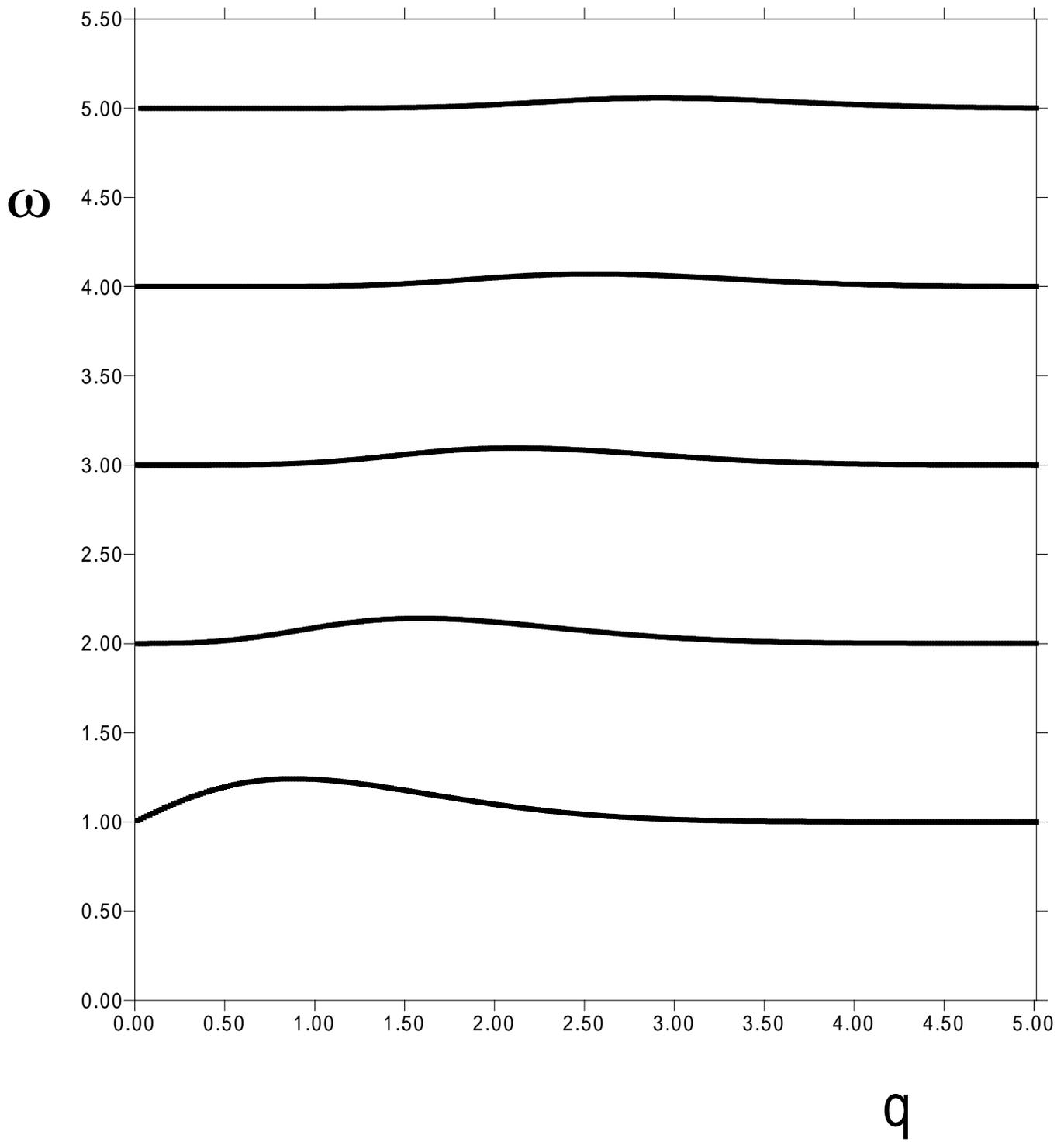
Figure 1: \\Collective modes for $\nu=1$, $r_{s}=1$.

\noindent  
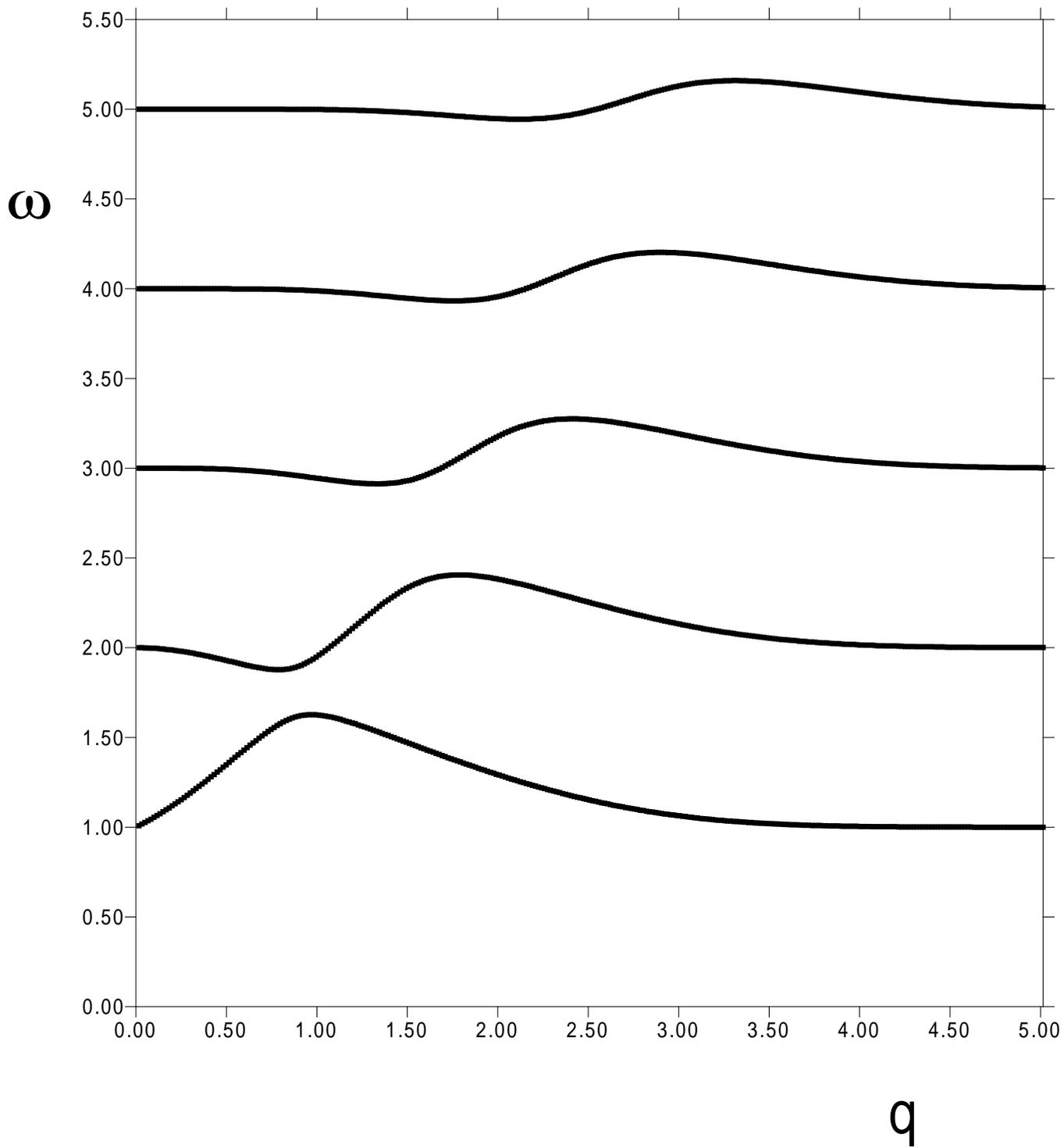
Figure 2: \\Collective modes for the filling $\nu=1$ given within the 
$p=1$ CF description ($\nu^{*}=-1$), $r_{s}=1$.

\noindent  
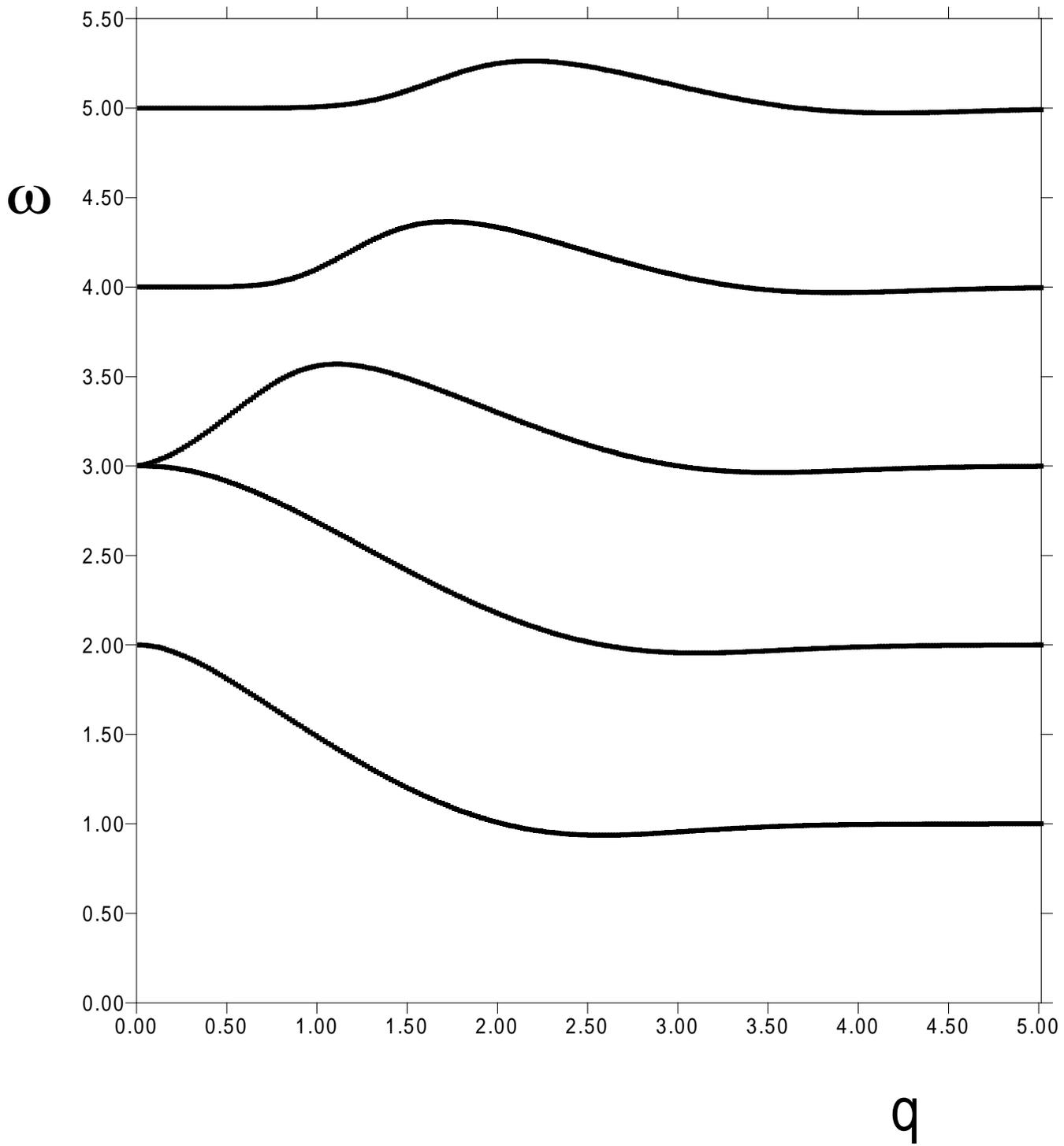
Figure 3: \\Collective modes for $\nu=1/3$, $r_{s}=1$.
  
\noindent
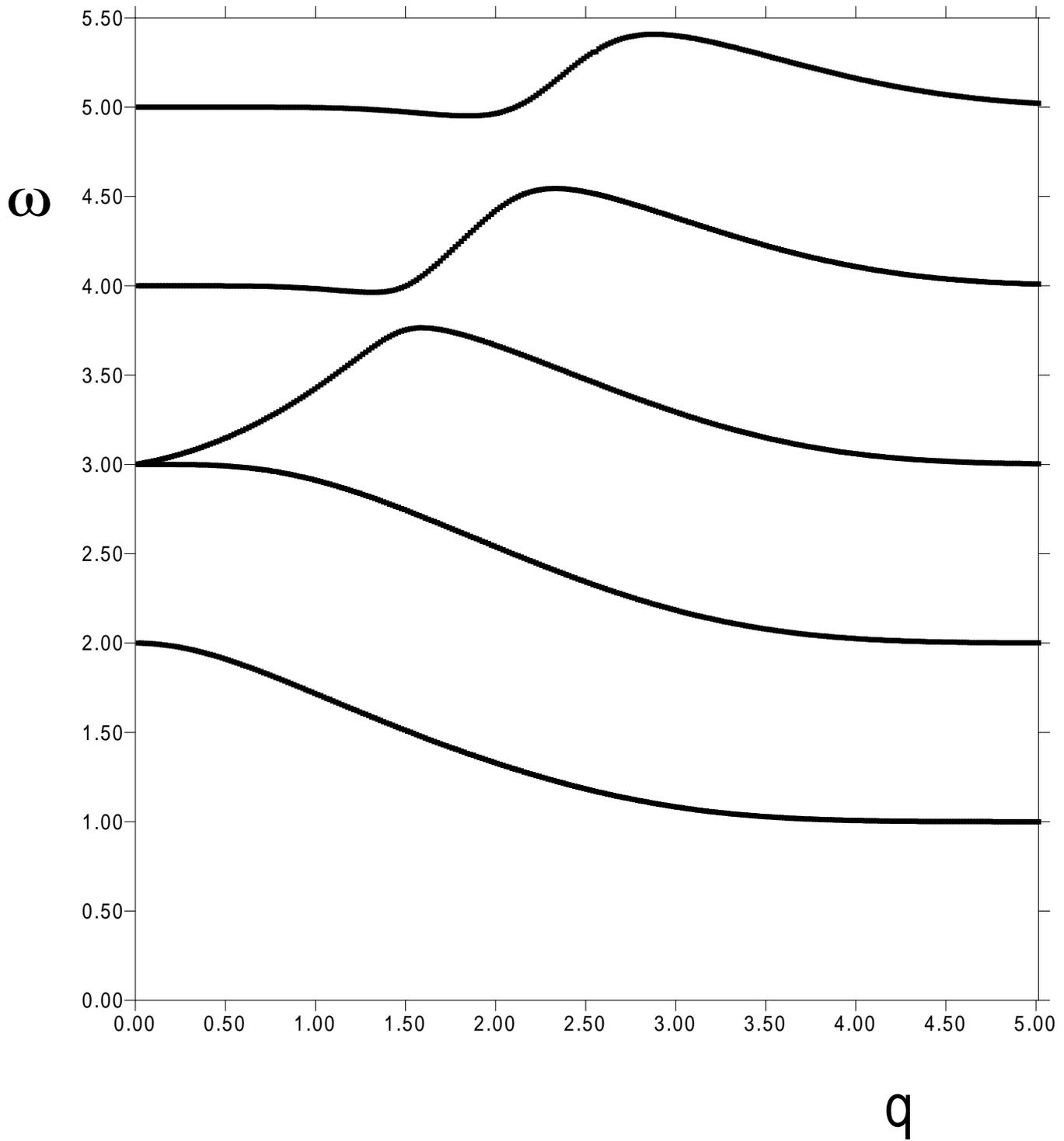
Figure 4: \\Collective modes for the filling $\nu=1/3$ given within the 
$p=2$ CF description ($\nu^{*}=-1$), $r_{s}=1$.

\noindent
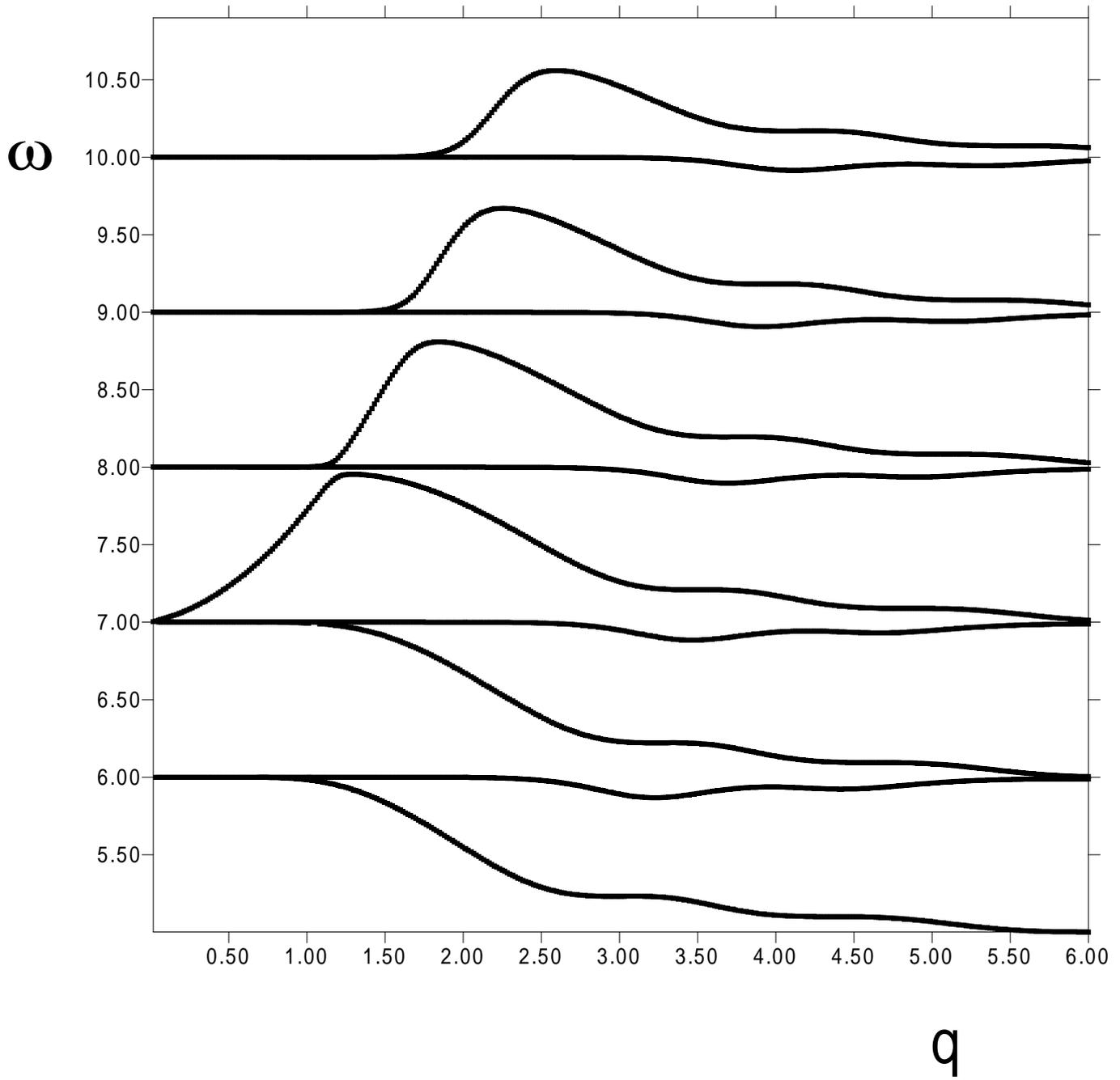
Figure 5: \\Collective modes for $\nu=3/7$, $r_{s}=1$.

\vspace{1cm}

\newpage
\begin{center}
\begin{tabular}{|c|c|c|}
\hline\hline
$\nu$&  H-F & exact\\ \hline 
$1$& $-0.627$ & $-0.627$\\
$2/3$& $-0.497$  & $-0.519$\\
$5/11$& $-0.406$ & $-0.451^{*}$ \\
$4/9$& $-0.402$ & $-0.447^{*}$ \\
$3/7$& $-0.396$   & $-0.443$\\
$2/5$& $-0.385$  &$-0.433$\\
$1/3$& $-0.362$& $-0.412$\\
$1/5$&$-0.280$  &$-0.328$\\
\hline\hline
\end{tabular} 
\end{center}
Table I: The Hartree-Fock and exact interaction energies 
(per particle) in respective units of 
$\frac{e^{2}}{\epsilon a_{0}^{ex}}$. 
The "exact" results are taken from Refs. \cite{Fano,Morf}
where the  results (in spherical systems) for few particles ($N\le 12$)
were extrapolated for infinite systems ($N\rightarrow\infty$).
Two numbers with stars
are obtained within the Jain CF approach \cite{Jain2}.
The "exact" $2/3$ result is 
found via particle-hole symmetry \cite{Fano2}.

\newpage
\vspace{0.5cm}

\begin{center}
\begin{tabular}{|c|c|c|c|c|c|c|c|c|}
\hline\hline
$\nu$&$\nu^{*}$&$p$&
\multicolumn{6}{|c|}{$\frac{\Delta E_{c}^{RPA}}{N}$} \\ 
\cline {4-9}
& & & \multicolumn{3}{|c|}{$r_{s}=10^{-4}$} &
\multicolumn{3}{|c|}{$r_{s}=10^{-5}$} \\
\cline {4-9}
& & & $k=10$ & $k=15$ & $k=20$ & $k=10$ & $k=15$ & $k=20$\\
\hline
$3/7$ & $3$ & $1$ & $-0.531$   & $-0.525$   &  $-0.520 $  & $-0.531$  & 
$-0.525$ & $-0.520$  \\ 
$2/5$ &$2$& $1$ & $-0.360$ & $-0.357$ & $-0.355$ & $-0.360$ & $-0.357$ & 
$-0.355$\\
$1/3$ &$1$& $1$ & $-0.247$ & $-0.247$ & $-0.246$ & $-0.247$ & $-0.247$ &
$-0.246$\\
$1/5$&$1$& $2$ & $-0.548$ & $-0.544$ & $-0.542$ & $-0.548$ & $-0.544$ & 
$-0.542$\\
$1$&$-1$&  $1$ &  $-0.230$ & $-0.230$ & $-0.230$&   $-0.230$ & $-0.230$ &
$-0.230$\\
$1/3$&$-1$& $2$ & $-0.585$ & $-0.581$ & $-0.580$ & $-0.585$ & $-0.581$ &
$-0.580$\\
$2/3$&$-2$& $1$ & $-0.358$ & $-0.354$ & $-0.352$ & $-0.358$ & $-0.354$ &
$-0.352$\\
\hline\hline
\end{tabular} 
\end{center}
Table II. The RPA correlation energies 
in units of $\frac{e^{2}}{\epsilon
a_{0}^{eff}}$.
For the $k=10$ case of $\nu=3/7$ the summation goes over $15$ modes.
For negative effective fillings we used instead
$\nu^{*}>0$ $p<0$ ($\nu <0$).


\begin{center}
\begin{tabular}{|c|c|c|c|c|c||c|}
\hline\hline
$\nu$& $\nu^{*}$ &$p$& $\frac{E^{H-F}}{N}$ & $\frac{\Delta E_{c}^{RPA}}{N}$ &
$\frac{E^{H-F}}{N}+\frac{\Delta E_{c}^{RPA}}{N}$ & exact diagonalization \\ \hline
$1$& &$0$ & $-0.627$ & $0$  & $-0.627$ &  $-0.627$\\  
$3/7$  & $3$ & $1$ & $-0.396$ & $-0.197$ & $-0.593$ & $-0.443$\\
$2/5$ &$2$ & $1$ & $-0.385$ & $-0.159$ &  $-0.544$ & $-0.433$\\
$1/3$ &$1$ & $1$ & $-0.362$ & $-0.142$ & $-0.504$ & $-0.412$\\
$1/5$&$1$ & $2$ & $-0.280$ & $-0.242$ & $-0.522$ & $-0.328$\\
  \hline\hline
\end{tabular} 
\end{center}
Table III. The interaction energies in respective 
units of $\frac{e^{2}}{\epsilon
a_{0}^{ex}}$ (note the change of units with respect to Table II). 
The RPA values in the fifth column are taken from
$k=20$ results of Table II.

\newpage
\begin{center}
\begin{tabular}{|c|c|c|c|c|c||c|}
\hline\hline
$\nu$& $\nu^{*}$ &$p$& $\frac{E^{H-F}}{N}$ & $\frac{\Delta E_{c}^{RPA}}{N}$ &
$\frac{E^{H-F}}{N}+\frac{\Delta E_{c}^{RPA}}{N}$ & exact diagonalization \\ \hline
$1$&$-1$ &$1$ & $-0.627$ & $-0.230$ & $-0.857$ & $-0.627$\\
$1/3$ &$-1$ & $2$ & $-0.362$ & $-0.335$ & $-0.697$ & $-0.412$\\
$2/3$&$-2$ & $1$ & $-0.497$ & $-0.203$ &  $-0.700$ & $-0.519$\\
  \hline\hline
\end{tabular} 
\end{center}
Table IV. The interaction energies in units of $\frac{e^{2}}{\epsilon
a_{0}^{ex}}$
obtained for negative effective fillings.

\end{document}